\documentstyle[12pt,epsf]{article}
 \newcommand {\bi} {\bibitem}
 \newcommand {\be} {\begin{equation}}
\newcommand {\bea} {\begin{eqnarray} \nonumber }
\newcommand {\ee} {\end{equation}}
\newcommand {\eea} {\end{eqnarray}}
 \newcommand {\eps} {\epsilon}
 \newcommand {\si} {\sigma}
\newcommand {\de} {\delta}

\newcommand {\for} {\ \ \ \mbox{for}\ \ }

\def \form#1 {eq. (\ref{#1}) }
\def \parziale#1#2  {{\partial {#1} \over \partial {#2}}}

\begin{document}

\title{Off-equilibrium fluctuation dissipation relation in binary mixtures}
\author{
Giorgio
Parisi\\
Dipartimento di Fisica, Universit\`a {\em La  Sapienza},\\ 
INFN Sezione di Roma I \\
Piazzale Aldo Moro, Rome 00185}
\maketitle

\begin{abstract}
In this note we present numerical simulations of binary mixtures.  We study the diffusion of 
particles and the response to an external driving force.  We find evidence for the validity of the 
Cugliandolo Kurchan off-equilibrium fluctuation dissipation relation.  The results are in agreement 
with the predictions of one step replica symmetry breaking and the dependance of the breakpoint 
parameter on the temperature coincides with that found in simple generalized spin glass models.
\end{abstract}

The behaviour of an Hamiltonian system (with dissipative dynamics) approaching equilibrium is well 
understood in a mean field approach for infinite range disordered systems \cite{CUKU,FM,BCKM}.  In 
this case we must distinguish an high and a low temperature region.  In the low temperature phase the 
correlation and response functions satisfy some simple relations derived by Cugliandolo and Kurchan
\cite{CUKU}.  In a previous note \cite{PAAGE1,PAAGE2} we have found some preliminary evidence for
the validity of these relations.  Here we perform a different and more accurate numerical experiment 
and we are able to study the temperature dependance of the phenomenon.

Let us define our notations.  We concentrate our attention on a quantity $A(t)$, which depends on 
the 
dynamical variables $x(t)$.  Later on we will make a precise choice of the function $A$.
We suppose that the system starts at time $t=0$ from a given initial condition and subsequently it 
is at a given temperature $T$.  If the initial configuration is at 
equilibrium at a temperature $T'>T$, we observe an off-equilibrium behaviour when we change the 
temperature.  In this note we will consider only the case $T'>>T$ (in particular we will study the 
case $T'=\infty$).

We can define a correlation function
\be
C(t,t_{w}) \equiv <A(t_{w}) A(t+t_{w})>
\ee
and the response function
\be
G(t,t_{w}) \equiv \frac{ \de A(t+t_{w})}{\de \eps(t_{w})}{\Biggr |}_{\eps=0},
\ee
where we are considering the evolution in presence of a time dependent Hamiltonian in which we have
added the term
\be
\int dt \eps(t) A(t).
\ee

The off-equilibrium fluctuation dissipation relation (OFDR)  states some among  the 
correlation functions and response function in the limit $t_{w}$ going to infinity.  The usual
equilibrium fluctuation  dissipation theorem (FDT)  tell us that
\be G(t)= - \beta \frac{d C(t)}{dt}, \ee
where
\be
G(t)=\lim_{t_w \to \infty} G(t,t_w), \ \ C(t)=\lim_{t_w \to \infty} C(t,t_w).
\ee

It is convenient to define the integrated response:
\be
R(t,t_{w})=\int_{0}^{t} d\tau G(\tau,t_{w}),\ \ R(t)=\lim_{t_w \to \infty} R(t,t_w),
\ee
which is the response of the system to a field acting for a time $t$.

The usual FDT relation is 
\be
R(t)= \beta (C(t)-C(0)).
\ee
The off-equilibrium fluctuation dissipation relation state that the 
response function and the correlation functions satisfy the following relations for large $t_w$:
\be
R(t,t_w)\approx \beta \int_{C(t,t_w)}^{C(0,t_{w})}X(C) dC.  \label{OFDR}
\ee
In other words
for large $t_{w}$ if we plot $R$ versus $C$ the data  collapse on the same universal curve
and the slope of that curve is $X(C)$.  The function $X(C)$ is system dependent and its form tells us 
 interesting information.

We must distinguish two regions:
\begin{itemize}
\item A short time region where $X(C)=1$ (the so called FDT region) and $C$ belongs to the interval
$I$
(i.e. $C_1<C<C_2$.).

 \item  A large time region (usually $t=O(t_w)$) where 
$C\notin I$ and $X(C)<1$ (the aging region) \cite{B,POLI}.
\end{itemize}

In the simplest non trivial case, i.e.  one step replica symmetry breaking \cite{mpv,parisibook2} , 
the function $X(C)$ is piecewise constant, i.e.
\bea
X(C)= m \for C \in I,\\
X(C)= 1 \for C \notin I \label{ONESTEP}.
\eea

In all known cases in which one step replica symmetry holds, the quantity $m$ vanishes linearly with 
the temperature at small temperature.  It often happens that $m=1$ at $T=T_{c}$ and $m(t)$ is 
roughly linear in the whole temperature range.  The relation \ref{OFDR} has been numerically 
verified in ref.  \cite{FRARIE}> The previous considerations are quite general and can be applied 
also to systems without quenched disorder
\cite{MPR}.   

If replica symmetry is broken at one step, the value of $m$ does not depend  on
the observable $A$ and the same value of $m$ should be obtained for all the observables.  In this 
case the OFDR has an highly predictive power because the value of $m$ may be measured by using quite 
different quantities.

The aim of this note is to show that for a binary mixture of spheres the function $X(C)$ is similar 
to predictions the one step formula (\ref{ONESTEP}) with a linear dependance of $m$ on the 
temperature.  It was already shown that in this model simple aging is well satisfied and some 
indications for the validity of the OFDR were already found \cite{PAAGE1,PAAGE2} by looking to the 
correlations of the stress energy tensor.  In this note we study a different observable and we find 
much more accurate results which confirm the previous findings.

We consider  a mixture of soft particles of different sizes.  Half of the particles 
are of type $A$, half of type $B$ and the interaction among the particle is given by the 
Hamiltonian:
\begin{equation}
H=\sum_{{i<k}} \left(\frac{(\si(i)+\si(k)}{|{\bf x}_{i}-{\bf x}_{k}|}\right)^{12},\label{HAMI}
\label{HAMILTONIAN}
\end{equation}
where the radius ($\si$) depends on the type of particles.  This model has been 
carefully studied  in the past \cite{HANSEN1,HANSEN2,HANSEN3,LAPA}.  It is known that a 
choice of the radius such that $\si_{B}/\si_{A}=1.2$ strongly inhibits crystallisation and the 
systems goes into a glassy phase when it is cooled.  Using the same conventions of the previous 
investigators we consider particles of average diameter $1$, more precisely we set
\begin{equation} 
{\si_{A}^{3}+ 2 (\si_{A}+\si_{B})^{3}+\si_{B}^{3}\over 4}=1.
\label{RAGGI}
\end{equation}
 
Due to the simple scaling behaviour of the potential, the thermodynamic quantities depend only on 
the quantity $T^{4}/ \rho$, $T$ and $\rho$ being respectively the temperature and the density 
(which we take equal to 1).  The model has been widely studied especially for this choice of the 
parameters.  It is usual to introduce the quantity $\Gamma \equiv \beta^{4}$.  The glass transition 
is known to happen around $\Gamma_c=1.45$ \cite{HANSEN2}.

Our simulation are done using a Monte Carlo algorithm.  We start by placing the particles at random 
and we quench the system by putting it at final temperature (i.e.  infinite cooling rate).  Each 
particle is shifted by a random amount at each step, and the size of the shift is fixed by the 
condition that the average acceptance rate of the proposal change is about .4.  Particles are 
placed 
in a cubic box with periodic boundary conditions.  In our simulations we have considered a 
relatively small number of particles, i.e.  $N=66$.  Previous studies have shown that such a small 
sample is quite adequate to show interesting off-equilibrium behaviour.

The main quantity on which we will concentrate our attention is the diffusion of the particles. i.e. 
\be
\Delta(t,t_{w})\equiv {\sum_{i=1,N}|{\bf x}_{i}(t_{w})-{\bf x}_{i}(t_{w}+t))|^{2} \over N}
\ee
The usual diffusion constant is given by $D=\lim_{t\to\infty}\Delta(t,t_{w})/t$.

The other quantity which  we measure is the response to a force.  We add at time $t_{w}$ the term
$\eps {\bf f} \cdot {\bf x}_{k}$, where $f$ is vector of squared length equal to 3 (in three 
dimensions) and we measure the response
\be
R(t_{w},t)={\partial {\bf f} \cdot {\bf x}_{k}(t_{w}+t)) \over \partial \eps}{\Biggr |}_{\eps=0}
\approx { {\bf f} \cdot {\bf x}_{k}(t_{w}+t)) \over  \eps}
\ee
for sufficiently small $\eps$.

 \begin{figure}[htbp]
\epsfxsize=400pt\epsffile[22 206 549 549]{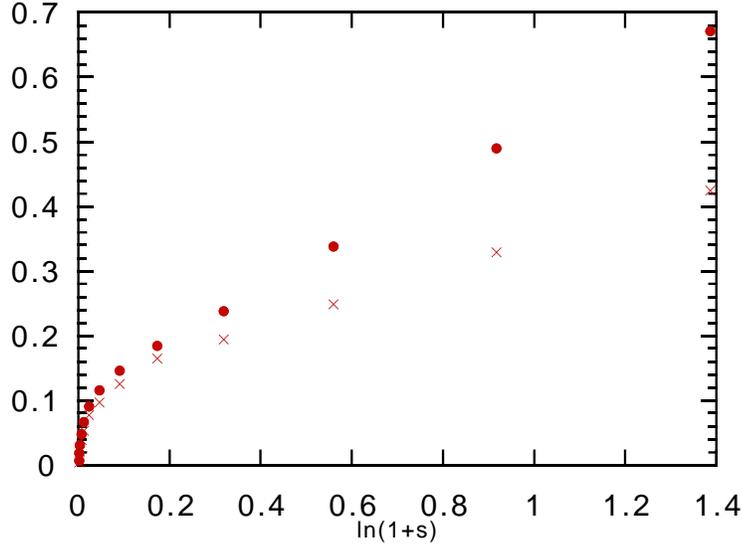}
\caption{ $\beta \Delta$ (points) and $R$ (crosses) as function the ratio $\ln(1+s)\equiv 
\ln(1+t/t_{w})$ at
$\Gamma=1.7$
for
$t_{w}=2048.$}
\end{figure}

The usual fluctuation theorem tells that at equilibrium $\beta \Delta=R$. This relation holds in
spite of the fact that  $\Delta(t,t_{w})$ is not the product of two observables one at time $t_w$,
the other at time $t_w+t$. However it can be written as 
\be
 \Delta(t,t_{w})\equiv {\sum_{i=1,N}{\bf x}_{i}(t_{w})^2+{\bf x}_{i}(t_{w}+t))^{2}
-2 {\bf x}_{i}(t_{w}) \cdot {\bf x}_{i}(t_{w}+t))} / {N}
\ee
and a detailed analysis \cite{CKP} shows that the fluctuation dissipation theorem is valid also in
this case.

 In the following we will look for the validity in 
the low temperature region of the generalized relation $\beta X(\Delta) \Delta=R$.  This relation 
can be valid only in the region where the diffusion constant $D$ is equal to zero.
Strictly speaking also in the glassy region $D\ne 0$, because diffusion may always happens by
interchanging two nearby particles ($D$ is different from zero also in a crystal); however if the
times are not too large the value of $D$ is so small in the glassy phase that this process may be
neglected in a first approximation.

\begin{figure}[htbp]
\epsfxsize=400pt\epsffile[22 206 549 549]{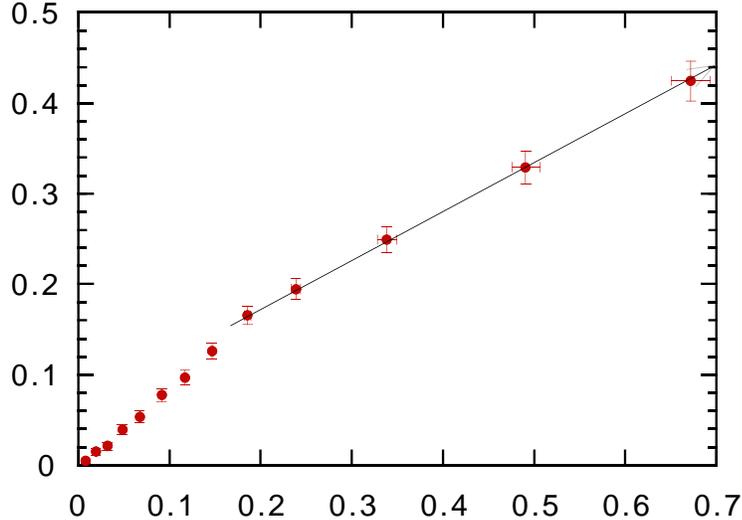}
\caption{ $R$ versus $\beta\Delta$ at $\Gamma=1.6$ for
$t_{w}=t_{w}=8192.$ and $t_{w}=2048$}
\end{figure}

We have done simulations for $N=66$.  For $N=66$ we average over 250 samples in absence of the 
force 
and on 1000 samples in presence of the force.  We have done simulations for $\Gamma=1.4,1.5 ...  
2.0\ $.  The dynamics was implemented using the Monte Carlo method, with an acceptance rate fixed 
around .4.  In order to decrease the error on the determination of $R$ we follow (as suggested in
\cite{CuKu97}) 
the method of computing in the same simulation the response function for different particles 
\cite{BMP}.  In other words we add to the Hamiltonian the term $\eps
\sum_{k=1,N}{\bf} f_{k} \cdot {\bf x}_{k}$ where the $f$ are random Gaussian vectors of
average squared length equal to 3.
The quantity $R$ can be computed as the average over $f$ of
\be
{\sum_{k=1,N} {\bf f}_{k} \cdot {\bf x}_{k}(t_{w}+t)) \over N \eps}.
\ee

The value of $\eps$ should be sufficient small in order to avoid non linear effects: we have done
extensive tests for $\eps=.1$ and $\eps=.2$.  We found that $\eps=.2$ is in the linear region, but 
we have followed the more conservative option $\eps=.1$.  We present the 
results for $t_{w}=2048$.  We have done also some simulations at $t_{w}=8192$, but we have not 
observed any systematic shift.

 In fig.1 we show the dependence of $\Delta$ and $R$ on the ratio $s\equiv 
t/t_{w}$ in the low temperature region, i.e.  at $\Gamma=1.7$ for $t_{w}=2048$.  They coincide in 
the small $s$ region, where FDT holds, but they differ for $s>.1$.

In fig.2 we show $R$ versus $\beta\Delta$ at 
$t_{w}=2048$, still at  $\Gamma=1.7$.   We see two linear regions with different slope as 
follows from the assumption of one step replica symmetry breaking.  The slope in the first region 
is 1, as expected form the FDT theorem, while tre slope in the second region is definitively smaller 
that 1.

Also the data at different temperatures for all values of $\Gamma\ge 1.5$ show a similar behaviour.  
The value of $R$ in the region where the FDT relation does not hold can be very well fitted by a 
linear function of $\Delta$ as can be seen in fig.  2.The region where a linear fit (with $m<1$)
is quite 
good correspond to $t/t_w>.2$.  The fitted value of $m\equiv \partial R/\partial (\beta \Delta)$ is 
displayed in fig.  3.  When $m$ becomes equal to 1, the fluctuation dissipation theorem holds in the 
whole region and this  happens at higher temperatures.  The straight line is the prediction 
of the approximation $m(T)=T/T_{c}$, using $\Gamma_{c}=1.45$ which seems to be rather good..

The value of $m$ we find at $\Gamma=1.8$ (i.e. $m=.33\pm .04$) is compatible with the value
$(m=.3\pm .1$) of ref. \cite{PAAGE2} extracted from the fluctuation of the stress energy tensor.
The method described in this note is much more accurate for two reasons:
\begin{itemize}
\item
The quantities which we consider becomes larger when we enter in the OFDR region:  they increase 
(not
decrease) as function of time.
\item
The correlation quantity we measure is an intensive quantity which becomes self averaging in the 
limit
of infinite volume.
\end{itemize}

 \begin{figure}[htbp]
\epsfxsize=400pt\epsffile[22 206 549 549]{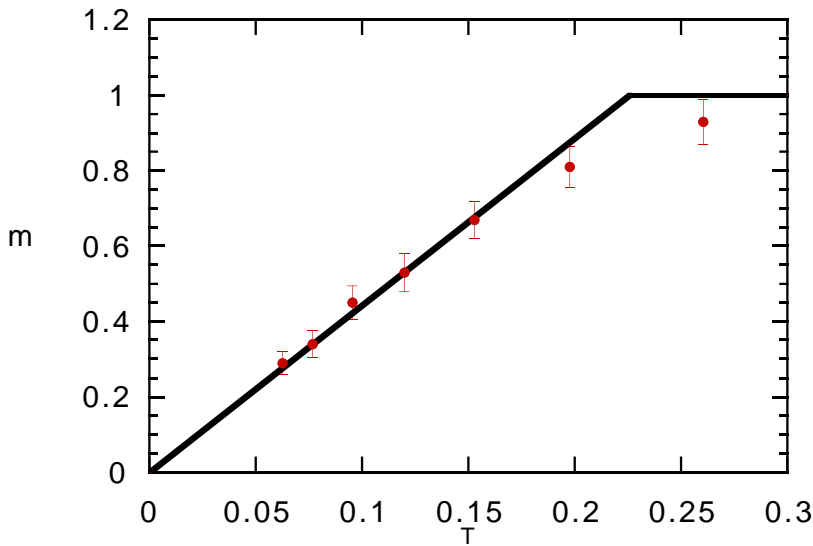}
\caption{ The quantity $m\equiv {\partial R\over \partial \beta \Delta}$ as $t_{w}=2048$ as 
function of the temperature. The straight line is 
the prediction of the approximation $m(T)=T/T_{c}$.}
\end{figure}

It is amusing to notice (as stressed to me by G. Ruocco) that the simple aging relation 
$\Delta(t_{w},t_{w})=cost$ for large $t_{w}$ implies that the particle move in average by a constant 
amount in each interval of time $2^{K}<t<2^{K+1}$.  If we assume that the movement in each time 
interval are uncorrelated, it follows that in the glassy phase $\Delta(t,t_{w})
\propto \ln(1+t/t_{w})$ for not too small $t/t_{w}$.  This is what happens outside the FDT region, 
(i.e. $t/t_{w}>.2$), as it can be seen from fig.  1.  In fig.  4 we
show the data for $t_{w}=1$, i.e.  the distance from the initial configuration.  The data seem to 
display a very nice logarithmic behaviour. 

All these results are in very good agreement with the theoretical expectations based on our knowledge 
extracted from the mean field theory for generalized spin glass models.  The approximation 
$m(T)=T/T_{c}$ seems to work with an embarrassing precision.  We can conclude that the ideas 
developed for generalized spin glasses have a much wider range of application than the models from 
which they have been extracted.  It likely that they reflect quite general properties of the phase 
space and therefore they can be applied in cases which are quite different from the original ones.

\begin{figure}[htbp]
\epsfxsize=400pt\epsffile[22 206 549 549]{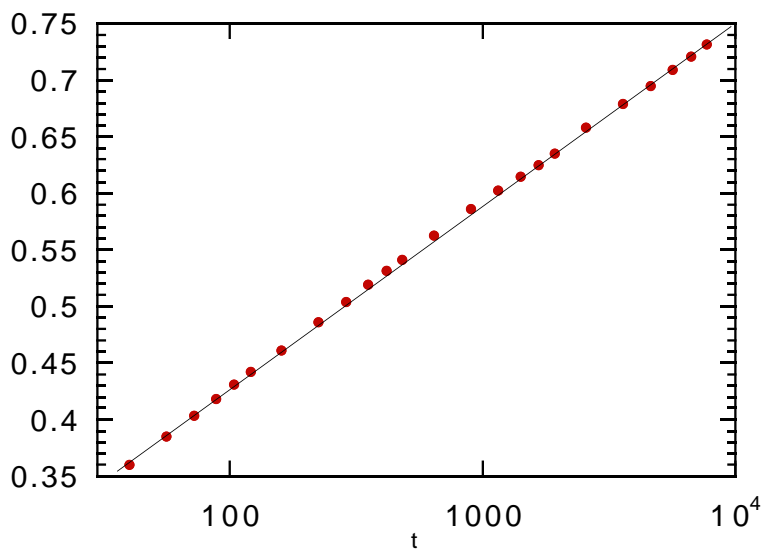}
\caption{ The distance from the initial configuration as function of the time in a logarithmic 
scale.}
\end{figure}

The most urgent theoretical task would be now to develop an analytic theory for glasses in the low 
temperature region from which one could compute the function $m(t)$.  This goal should not be out of 
reach: a first step in this direction can be found in \cite{MEPA}

\section* {Acknowledgments} I thank  L.
Cugliandolo, S.  Franz, J.  Kurchan and G.  Ruocco for many useful discussions and suggestions.


\begin{thebibliography}{99}
 
\bi{CUKU} L.  F.  Cugliandolo and J.Kurchan, Phys.  Rev.  Lett.  {\bf 71}, 
1 (1993).

\bi{FM} S. Franz and M. M\'ezard {\it On mean-field glassy
dynamics out of equilibrium},  cond-mat 9403004. 

\bi{BCKM} J.-P.  Bouchaud, L.  Cugliandolo, J.  Kurchan, Marc M\'ezard, cond-mat 9511042.

\bibitem{PAAGE1} G.  Parisi, {\sl Short time aging in binary glasses}, cond-mat 9701015.

\bibitem{PAAGE2} G.  Parisi, {\sl Numerical indications for the existence of a thermodynamic
     transition in binary glasses}, cond-mat 9701100.

\bi{B} J.-P. Bouchaud; J. Phys. France {\bf 2} 1705, (1992). 

\bi{POLI} L. C. E. Struik; {\it Physical aging in amorphous polymers and other
 materials} (Elsevier, Houston 1978).

\bibitem{mpv} M.M\'ezard, G.Parisi and M.A.Virasoro, {\sl Spin glass theory and 
beyond}, World Scientific (Singapore 1987).
\bibitem{parisibook2} G.Parisi, {\sl Field Theory, Disorder and
Simulations}, World Scientific, (Singapore 1992).

\bi{FRARIE} S. Franz and H. Rieger Phys.  J. Stat. Phys.  {\bf 79} 749 (1995).

\bi{MPR}E.  Marinari, G.  Parisi and F.  Ritort,
J.  Phys.  A: Math.  Gen.  {\bf 27} 7615 (1994).

\bi{HANSEN1}B.Bernu, J.-P. Hansen, Y. Hitawari and G. Pastore, Phys. Rev. {\bf A36}
 4891 (1987).

\bi{HANSEN2}J.-L. Barrat, J-N. Roux and J.-P. Hansen, Chem. Phys. {\bf 149}, 197
(1990).

\bi{HANSEN3}J.-P. Hansen and S. Yip, Trans. Theory and Stat. Phys. {\bf 24}, 1149
(1995).

\bibitem{LAPA} D.  Lancaster and G.  Parisi, {\sl A Study of Activated Processes in Soft Sphere 
Glass}, cond-mat 9701045.

\bibitem{CKP} L.  Cugliandolo and J.  Kurchan and G.  Parisi, {\sl Off equilibrium dynamics and 
aging in unfrustrated systems}, cond-mat/9406053, J.  Phys.  I (France) {\bf 4} 1691 (1994).

\bibitem{CuKu97} L. Cugliandolo and J. Kurchan, private communication.

\bibitem{BMP} A. Billoire, E. Marinari, G. Parisi,  Phys. Lett. {\bf 162B} (1985) 160.
     
\bi{MEPA} M.M\'ezard and G.Parisi, {\sl A tentative Replica Study of the Glass Transition}, 
cond-mat/9602002.

\end{thebibliography}
\end{document}